\documentclass[12pt]{article}
\usepackage{fullpage,authblk,cite}
\usepackage{amsmath,amsthm,graphicx,amssymb,verbatim,listings}
\usepackage{wasysym,color,enumitem,mathcomp}
\usepackage[T1]{fontenc}     % needed for the guillemets
\usepackage{lmodern, tipa}   % needed to fix suboptimal font rendering
\usepackage[ pdftex, plainpages = false, pdfpagelabels,
                 pdfpagelayout = useoutlines,
                 bookmarks,
                 bookmarksopen = true,
                 bookmarksnumbered = true,
                 breaklinks = true,
                 linktocpage=all,
                 pagebackref=false,
                 colorlinks = true,
                 linkcolor = Blue,
                 urlcolor  = blue,
                 citecolor = BrickRed,
                 anchorcolor = green,
                 hyperindex = true,
                 hyperfigures
                 ]{hyperref}
\usepackage[usenames, dvipsnames]{xcolor}
\usepackage{xifthen} % provides \isempty test

\newcommand{\tr}{\hbox{tr}}

\newcommand{\arxiv}[2][]{\ifthenelse{\isempty{#1}}{\href{http://arxiv.org/abs/#2}{{\tt arXiv:\allowbreak{}#2}}} {\href{http://arxiv.org/abs/#2}{{\tt arXiv:\allowbreak{}#2 [#1]}}}}

\newcommand{\booktitle}{\textsl}
\newcommand{\hrefdoi}[2]{\href{https://dx.doi.org/#1}{#2}}

\begin{document}
\title{The Status of the Bayes Rule in QBism}
\author[$\dag$]{Blake C.\ Stacey}
\affil[$\dag$]{\small{\href{http://www.physics.umb.edu/Research/QBism}{QBism Group}, Physics Department, University of
    Massachusetts Boston\protect\\ 100 Morrissey Boulevard, Boston MA 02125, USA}}

\date{\small\today}

\maketitle

\begin{abstract}
I gamely try to disentangle ideas that have been confused with one another.  
\end{abstract}

The term ``Bayesian'' covers a broad territory and subsumes many
disputes~\cite{Good:1983}. For one pertinent example, it is possible
to be ``Bayesian'' --- to accept that probabilities are expressions of
knowledge, information or beliefs --- and reject the idea that the
Bayes conditionalization rule is the only means by which probabilities
can be modified over time. This is true of the interpretation of
probability that is part of QBism~\cite{Fuchs:2012}. As Fuchs and I
wrote in an earlier commentary, ``QBism subscribes to a school of
personalist probability in which an agent need not always update by
the Bayes conditioning rule, or even expect that she
will''~\cite{Fuchs:2020}. QBists are not original in this regard, as
Fuchs and Schack's bibliography~\cite{Fuchs:2012} makes clear. In this
tradition, one first derives the rules of probability theory from the
condition that an agent's beliefs should relate to each other
consistently. The simplest way to do so is typically a Dutch-book
argument. The rules of probability theory are normative constraints
with which an agent should strive to comply, and they constrain the
agent's beliefs \emph{at a single time.} Only in a later stage of the
argument and by invoking additional assumptions can we conclude
anything about the best way for an agent to \emph{change} her beliefs
over time.

We've noticed people getting this wrong about QBism, sometimes in
conversation but also in print, even in writings that are in other
ways rather insightful. This short note is to set the record straight,
so that people who want to be unhappy about QBism can be unhappy with
what it actually is. In addition, this will provide an opportunity to
look forward a bit and consider an active research problem.

To be explicit, let us say that Alice contemplates two events, $E$
that might happen on Monday and $F$ that might happen on
Tuesday. Before the possible time of either event, say on Sunday,
Alice codifies her beliefs into gambling commitments: $P_0(E)$,
$P_0(F)$ and the conditional probability $P_0(F|E)$. The subscript 0
indicates the time at which Alice commits to these probabilities. One
day later, the event $E$ occurs. Alice takes the opportunity to
reformulate her gambling commitments, picking a new value $P_1(F)$. Is
she \emph{obligated} to set
\begin{equation}
  P_1(F) = P_0(F|E)
\end{equation}
just because $E$ has transpired? This is a natural-looking choice and
doubtless justifiable in many circumstances, but it hardly appears
mandatory in general. Perhaps other events have taken place along with
$E$ that were so calamitous that Alice feels driven to reevaluate what
she thinks about the world much more broadly. (Frank Ramsey raised
this possibility in 1926~\cite{Ramsey:1931}; see
Zabell~\cite{Zabell:1991} and Misak~\cite{Misak:2020} for some
commentary.)

For one example of the literature getting this at least a little
wrong, consider the following~\cite{Glick:2019}.
\begin{quote}
  First, on the classic subjective Bayesian approach which inspires
  QBism, there are only coherence constraints---i.e., conformity to
  the probability axioms and Bayesian updating.
\end{quote}
Or, more explicitly, take the following passage in the opening of a
chapter posted as a draft to the PhilSci archive~\cite{French:2021}.
\begin{quote}
Crucially, the wave-function should be understood solely in epistemic
terms, as representing not some state of a physical system but rather
that of the agent with regard to their possible future experiences. It
does this by encoding the agent's coherent degree of belief in each of
certain alternative experiences that result from some act they
perform, such as those associated with the outcomes of a measurement
procedure. These beliefs are to be updated according to Bayes' Theorem:
\begin{displaymath}
P (A|B) = P(B|A) P(A)/P(B)
\end{displaymath}
where $A$ and $B$ are events, $P(A|B)$ is the conditional or
`posterior' probability of $A$ given $B$, $P(B|A)$ is the likelihood
of $A$ with $B$ fixed, and $P(A)$ and $P(B)$ are the `prior'
probabilities of the respective events.
\end{quote}
As the discussion above should make clear, the last sentence is not
quite accurate. (I've corresponded with French about this, and the
text will be corrected in the final version. I preserve it here, not
out of unkindness, but because it provides a crisp illustration of a
fallacy that I have more often heard stated implicitly or obscurely.)
To a QBist, wavefunction collapse is \emph{analogous} to an invocation
of the Bayes updating rule, in that it is a change of expectations due
to new experiences, but it is not an \emph{instance of} the Bayes
updating rule.

There is an interesting technical question in how closely
quantum-state update can be made to resemble the Bayes rule
algebraically, and under what circumstances. This question in fact
predates QBism proper~\cite{Fuchs:2002}. A general lesson about it is
that getting any rule that resembles the standard Bayes form is itself
dependent upon one's prior~\cite{Fuchs:2009}.

This is also a good opportunity to distinguish between the Bayes
updating rule and the \emph{Law of Total Probability.} Suppose that on
Sunday (day 0), Alice contemplates a whole set of possible happenings
on Monday, a mutually exclusive and exhaustive set of events
$\{E_i\}$. Again, let $F$ be an event which might happen on
Tuesday. If Alice is to be self-consistent in assembling her mesh of
beliefs, she ought to commit to gambling in accord with the relation
\begin{equation}
  P_0(F) = \sum_i P_0(F|E_i) P_0(E_i) \, .
\end{equation}
There is no \emph{updating} here:\ All of the time-index subscripts
are the same.

The warm sun around which most QBist technical work currently revolves
is the observation that the Born rule can be understood as a
counterpart for the LTP, operative in the case where the intermediate
measurement is not just ignored, but as Asher Peres would say,
\emph{unperformed}~\cite{Peres:1978}. That is, we can take the Born rule,
\begin{equation}
  P_0(F) = \tr(\rho F) \, ,
\end{equation}
and write it in entirely probabilistic terms:
\begin{equation}
  P_0(F) = \sum_j P_0(F|E_j) \sum_i [\Phi]_{ji} P_0(E_i) \, ,
\end{equation}
where the fact that the ``Born matrix'' $\Phi$ cannot be made to equal
the identity is a structural feature of quantum
theory~\cite{DeBrota:2020}. Again, there is no updating here:\ Every
time-index subscript is 0. Thus, statements like ``The central idea of
QBism is that the Born's rule is a quantum version of the Bayes' rule
for statistical inference''~\cite{D'Ariano:2016} also miss the mark,
in a subtle but significant way.

The QBist efforts toward reconstructing quantum theory have not
invoked the Bayes update rule, other than to shore up an auxiliary
postulate. This ``reciprocity'' axiom was, in Fuchs and Schack's
enumeration from a decade ago, Assumption 5 out of 7 --- so, a fair
way down the line~\cite{Fuchs:2011}. See also my book chapter with
Fuchs~\cite{Fuchs:2016} for a more conversational introduction to the
topic. This way of justifying the assumption never sat right. Among
other reasons, it gave an outsized importance to the state of maximal
ignorance (the flat probability distribution), an importance that was
hard to motivate. The appeal to an update rule also fit poorly with
the rest of the argument, which used only consistency conditions
between probabilities at a single time. Accordingly, we spent
considerable time trying to find alternate
justifications~\cite{Appleby:2017}, and lately we have taken to an
approach that makes the ``reciprocity'' condition a theorem rather
than a postulate~\cite{DeBrota:2021}. Overall, then, the only
appearance of the Bayes rule in the mainline of QBist work started off
marginal and has only been pushed further aside since.


\begin{thebibliography}{99}
\bibitem{Good:1983} I.\ J.\ Good, ``46,656 varieties of Bayesianism.''
  In \booktitle{Good Thinking:\ The Foundations of Probability and its
    Applications} (University of Minnesota Press, 1983).

\bibitem{Fuchs:2012} C.\ A.\ Fuchs and R.\ Schack,
  ``\hrefdoi{10.1007/978-3-642-21329-8_15}{Bayesian Conditioning, the
  Reflection Principle, and Quantum Decoherence}.'' In
  \booktitle{Probability in Physics,} edited by Y.\ Ben-Menahem and
  M.\ Hemmo. (Springer, 2012.) \arxiv{1103.5950}.

  \bibitem{Fuchs:2020} C.\ A.\ Fuchs and B.\ C.\ Stacey, ``QBians Do
    Not Exist,'' \arxiv{2012.14375} (2020).

  \bibitem{Ramsey:1931} F.\ P.\ Ramsey, ``Truth and probability.'' In
    \booktitle{Foundations of Mathematics and Other Logical Essays,}
    edited by R.\ B.\ Braithwaite. (Routledge and Kegan Paul, 1931.)

  \bibitem{Zabell:1991} S.\ L.\ Zabell,
    ``\hrefdoi{10.1111/j.1755-2567.1991.tb00838.x}{Ramsey, truth, and
    probability},'' \booktitle{Theoria} \textbf{57} (1991), 211--38.

  \bibitem{Misak:2020} C.\ Misak, \booktitle{Frank Ramsey:\ A Sheer
    Excess of Powers} (Oxford University Press, 2020).

  \bibitem{Glick:2019} D.\ Glick and F.\ J.\ Boge,
    ``\hrefdoi{10.1007/s10670-019-00163-w}{Is the Reality Criterion
    Analytic?},'' \booktitle{Erkenntis} (2019), 1--7, \arxiv{1909.11893}.
    
  \bibitem{French:2021} S.\ French, ``Putting Some Flesh on the
    Participant in Participatory Realism,''
    \url{http://philsci-archive.pitt.edu/19354/} (2021).
    
  \bibitem{Fuchs:2002} C.\ A.\ Fuchs, ``Quantum mechanics as quantum
    information (and only a little more),'' \arxiv{quant-ph/0205039}
    (2002).

  \bibitem{Fuchs:2009} C.\ A.\ Fuchs and R.\ Schack, ``Priors in
    Quantum Bayesian inference,'' \arxiv{0906.1714} (2009).

  \bibitem{Peres:1978} A.\ Peres,
    ``\hrefdoi{10.1119/1.11393}{Unperformed Experiments Have No
    Results},'' \booktitle{American Journal of Physics} \textbf{46}
    (1978), 745.

  \bibitem{DeBrota:2020} J.\ B.\ DeBrota, C.\ A.\ Fuchs and
    B.\ C.\ Stacey,
    ``\hrefdoi{10.1103/PhysRevResearch.2.013074}{Symmetric
      Informationally Complete Measurements Identify the Irreducible
      Difference between Classical and Quantum Systems},''
    \booktitle{Physical Review Research} \textbf{2} (2020), 013074,
    \arxiv{1805.08721}.

  \bibitem{D'Ariano:2016} G.\ M.\ D'Ariano and A.\ Khrennikov,
    ``\hrefdoi{10.1098/rsta.2015.0244}{Preface of the special issue
    quantum foundations: information approach},''
    \booktitle{Philosophical Transactions of the Royal Society A}
    \textbf{374} (2016), 20150244.
    
  \bibitem{Fuchs:2011} C.\ A.\ Fuchs and R.\ Schack,
    ``\hrefdoi{10.1007/s10701-009-9404-8}{A Quantum-Bayesian Route to
    Quantum State Space},'' \booktitle{Foundations of Physics}
    \textbf{41} (2011), 345--56, \arxiv{0912.4252}.

  \bibitem{Fuchs:2016} C.\ A.\ Fuchs and B.\ C.\ Stacey, ``Some
    Negative Remarks on Operational Approaches to Quantum Theory.''
    In \booktitle{Quantum Theory:\ Informational Foundations and
      Foils,} edited by G.\ Chiribella and
    R.\ W.\ Spekkens. (Springer, 2016.)  \arxiv{1401.7254}.

  \bibitem{Appleby:2017} M.\ Appleby, C.\ A.\ Fuchs, B.\ C.\ Stacey
    and H.\ Zhu, ``\hrefdoi{10.1140/epjd/e2017-80024-y}{Introducing
      the Qplex:\ A novel arena for quantum theory},''
    \booktitle{European Physical Journal D} \textbf{71} (2017), 197,
    \arxiv{1612.03234}.

  \bibitem{DeBrota:2021} J.\ B.\ DeBrota, C.\ A.\ Fuchs,
    J.\ L.\ Pienaar and B.\ C.\ Stacey,
    ``\hrefdoi{10.1103/PhysRevA.104.022207}{Born's rule as a quantum
      extension of Bayesian coherence},'' \booktitle{Physical Review
      A} \textbf{104} (2021), 022207, \arxiv{2012.14397}.

\end{thebibliography}
\end{document}